\documentclass[lettersize,journal]{IEEEtran}

\usepackage{cite}
\usepackage{amsmath,amssymb,amsfonts}
\usepackage{graphicx}
\usepackage{textcomp}
\usepackage{xcolor}
\usepackage{authblk}

\usepackage{booktabs}
\usepackage{epsfig}
\usepackage{latexsym}
\usepackage{multirow}
\usepackage{stfloats}
\usepackage{epstopdf}
\usepackage{color}  
\usepackage{tabularx} 
\usepackage{enumerate}
\usepackage{array}
\graphicspath{{./Figures/}}
\usepackage{color}
\usepackage{bbm}
\usepackage{caption}
\usepackage{bm}
\usepackage[tight,footnotesize]{subfigure}
\usepackage{balance}
\usepackage{mathrsfs}
\usepackage{verbatim}
\allowdisplaybreaks[4]
\usepackage{dsfont}
\usepackage{verbatim}
\usepackage{tikz}
\usepackage{setspace}
\usepackage{diagbox}
\usepackage[framemethod=tikz]{mdframed}
\usepackage{multicol}
\usepackage{environ}
\usepackage{tikz}
\usepackage{stfloats}
\usepackage{algpseudocode}
\usepackage{graphics}
\usepackage{epsfig}
\usepackage{amsthm}
\usepackage{authblk}
\usepackage{enumitem}
\usepackage{makecell}
\newcommand{\rev}[1]{\textcolor{black}{{#1}}}

\ifCLASSOPTIONcompsoc
\usepackage[caption=false, font=normalsize, labelfont=sf, textfont=sf]{subfig}
\else
\usepackage[caption=false, font=footnotesize]{subfig}

\usepackage{bm}
\def\BibTeX{{\rm B\kern-.05em{\sc i\kern-.025em b}\kern-.08em
    T\kern-.1667em\lower.7ex\hbox{E}\kern-.125emX}}
\begin{document}

\title{Far- and Near-Field Channel Measurements and Characterization in the Terahertz Band Using a Virtual Antenna Array}

\author{Yiqin~Wang,
Shu~Sun,
and~Chong~Han,~\IEEEmembership{Senior~Member,~IEEE}%
\thanks{Yiqin Wang is with Terahertz Wireless Communications (TWC) Laboratory, Shanghai Jiao Tong University, China (Email: wangyiqin@sjtu.edu.cn).}%
\thanks{Shu Sun is with Department of Electronic Engineering and the Cooperative Medianet Innovation Center (CMIC), Shanghai Jiao Tong University, China (Email: shusun@sjtu.edu.cn).}%
\thanks{Chong Han is with Terahertz Wireless Communications (TWC) Laboratory, also with Department of Electronic Engineering and the Cooperative Medianet Innovation Center (CMIC), Shanghai Jiao Tong University, China (Email: chong.han@sjtu.edu.cn).}}

\maketitle
\begin{abstract}

Extremely large-scale antenna array (ELAA) technologies consisting of ultra-massive multiple-input-multiple-output (UM-MIMO) or reconfigurable intelligent surfaces (RISs), are emerging to meet the demand of wireless systems in sixth-generation and beyond communications for enhanced coverage and extreme data rates up to Terabits per second.
For ELAA operating at Terahertz (THz) frequencies, the Rayleigh distance expands, and users are likely to be located in both far-field (FF) and near-field (NF) regions. On one hand, new features like NF propagation and spatial non-stationarity need to be characterized. On the other hand, the transition of properties near the FF and NF boundary is worth exploring.
In this paper, a complete experimental analysis of far- and near-field channel characteristics using a THz virtual antenna array is provided based on measurement of the multi-input-single-output channel with the virtual uniform planar array (UPA) structure of at most 4096 elements. In particular, non-linear phase change is observed in the NF, and the Rayleigh criterion regarding the maximum phase error is verified. Then, \rev{a new cross-field path loss model is proposed, which characterizes the power change at antenna elements in the UPA and is compatible with both FF and NF cases.}

\end{abstract} 
\section{Introduction}


While the lower part of the millimeter-wave (mmWave) frequency band (30-100~GHz) have been officially adopted in fifth-generation (5G) communications, sixth-generation (6G) is expected to explore higher frequency bands. Among candidate spectrum bands, the Terahertz (THz) band (0.1-10~THz, including the higher part of the mmWave band up to 300~GHz) is anticipated to address the spectrum scarcity and capacity limitations of current wireless systems. However, THz wave peculiarities, such as the narrow beam and strong path loss, constrain the transmission distance and signal coverage. In light of enormous spatial multiplexing and beamforming gain, ultra-massive multiple-input-multiple-output (UM-MIMO) and reconfigurable intelligent surfaces (RISs) that form extremely large-scale antenna array (ELAA) systems become a special interest for THz as they effectively increase the communication range and enhance the capacity in THz wireless communications~\cite{akyildiz2016realizing}.

Especially, thanks to the small wavelength of THz waves, electrically large-scale arrays with reasonable physical dimension becomes realistic.
However, for ELAA equipped with hundreds or thousands of antennas and high THz frequencies, features like near-field (NF) propagation and spatial non-stationarity arise and bring new challenges to channel modeling and characterization~\cite{cui2022near,channel_tutorial}.
In wireless communications, the electromagnetic (EM) radiation field is divided into the far-field (FF) region and the radiative NF region\footnote{The near-field region is subdivided into reactive NF and radiative NF. Since EM waves in the reactive NF are induced fields and do not set off from the antenna, we only refer to the radiative NF in this article.}, typically by the boundary called Rayleigh distance~\cite{selvan2017fraunhofer}, which is proportional to the square of array aperture and the carrier frequency. In THz ELAA systems, the Rayleigh distance increases as compared with lower frequency bands, and users are likely to be located in the NF region.
First, the planar wavefront model fails in the NF region, where the phase value along antenna elements cannot be approximated by a linear change.
Besides, in the NF region within the Rayleigh distance, the Friis' free-space path loss (FSPL) model is also no longer accurate~\cite{cui2022near,xiao2021near,kaurav2019determination}.

Till date, spherical wave propagation~\cite{sun2023how,zhou2015spherical,chen2021hybrid,li2022analytical,cui2022channel,xie2023near,bacci2023spherical,chen2023beam,chen2024triple} and non-stationarity~\cite{wu2015non,ali2019linear,cheng2019adaptive,han2020channel} have been analytically modeled, which are used to improve beamforming technologies on large-scale antenna arrays.
Nevertheless, experimental characterization of NF effects is rare~\cite{payami2012channel,martinez2014towards,yuan2023on}, let alone in the THz band.
To fill this research gap, in this paper, we provide a complete analysis of hybrid far- and near-field channel characteristics in the THz band based on measurement using a virtual antenna array (VAA). Specifically, we carry out wideband THz channel measurements of the multi-input-single-output (MISO) system with the uniform planar array (UPA) structure. Channels from 256, 1024, and 4096 elements in the UPA are investigated. \rev{In the static environment, VAA is free of mutual coupling and represents the ideal case of the actual antenna array.}
In light of the measurement results, we elaborate the variation of properties, i.e., path loss and phase, of the \rev{line-of-sight (LoS)} ray across elements in the UPA, in both FF and NF cases. First,\rev{a new cross-field path loss model is proposed, which characterizes the power change at antenna elements in the UPA and is compatible with both FF and NF cases.} Then, the Rayleigh criterion, which distinguishes NF and FF in terms of the maximum phase error, is verified.

The remainder of this paper is organized as follows.
The wideband THz virtual antenna array measurement is introduced in Sec.~\ref{section: campaign}. We then analyze channel measurement results in Sec.~\ref{section: results}. Finally, the paper is concluded in Sec.~\ref{section: conclusion}.
\section{Channel Measurement Campaign} \label{section: campaign}

In this section, we introduce the \rev{vector network analyzer (VNA)}-based THz MISO channel measurement platform, and describe the far-and-near-field measurement conducted in a small anechoic chamber wrapped by absorbing materials.

\begin{figure}
    \centering
    \subfigure[The diagram of the MISO measuring system.]{
    \includegraphics[width=\linewidth]{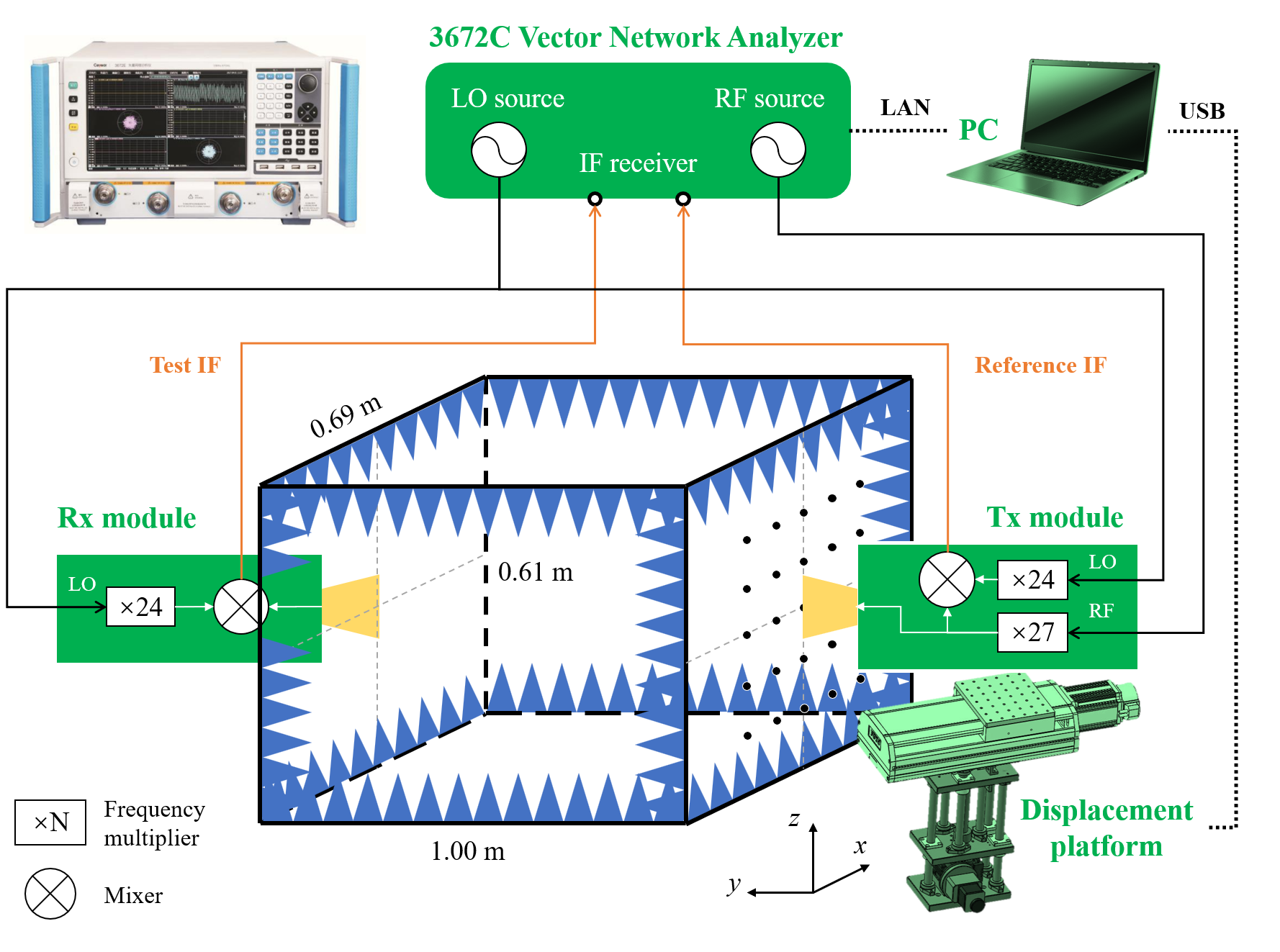}
    }
    \\
    \subfigure[The measurement scenario.]{
    \includegraphics[width=\linewidth]{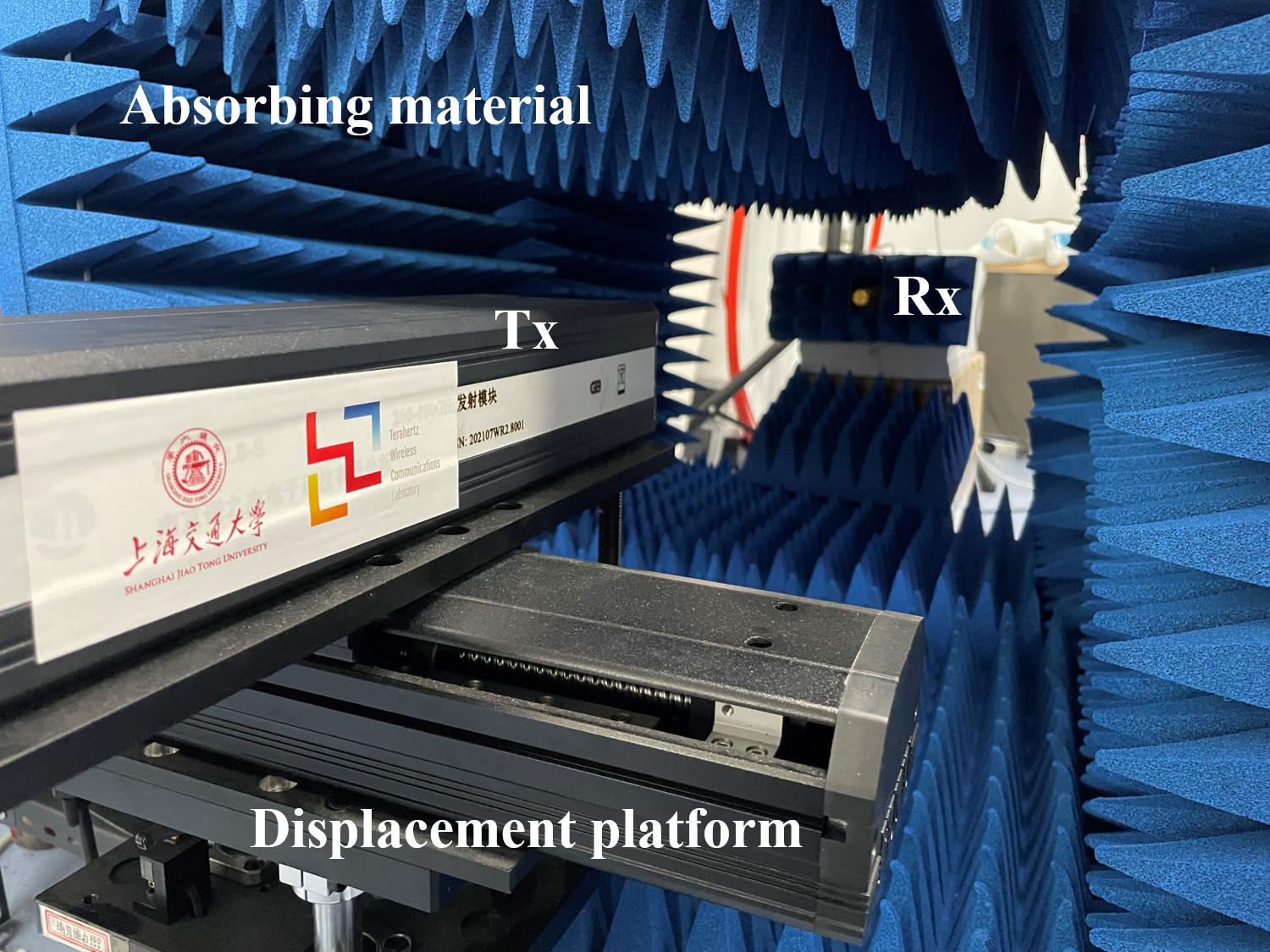}
    }
    \caption{\rev{Overview of the MISO measurement.}}
    \label{fig:platform}
\end{figure}

\subsection{Channel Measurement System}
The THz MISO channel measurement system is composed of three parts, as shown in Fig.~\ref{fig:platform}(a), including the computer (PC) as the control platform, the displacement platform, and the measuring platform.

The measuring platform consists of THz transmitter (Tx) and receiver (Rx) modules and the Ceyear 3672C VNA.
The VNA generates radio frequency (RF) and local oscillator (LO) sources.
The RF signal is multiplied by 27 to reach the carrier frequencies. The LO signal is multiplied by 24. As designed, the mixed intermediate frequency (IF) signal has a frequency of 7.6~MHz.
Two IF signals at Tx and Rx modules, i.e., the reference IF signal and the test IF signal, are sent back to the VNA, and the transfer function of the device under test (DUT) is calculated as the ratio of the two frequency responses. The DUT contains not only the wireless channel, but also the device, cables, and waveguides. To eliminate their influence, the system calibration procedure is conducted, which is described in detail in our previous works~\cite{wang2022thz, li2022channel}.

In this measurement, we investigate the THz frequency band ranging from 260~GHz to 320~GHz, which covers a substantially large bandwidth of 60~GHz. As a result, the time and space resolution is 16.7~ps and 5~mm, respectively.
The frequency sweeping interval is 60~MHz, resulting in 1001 sweeping points at each Tx-Rx position. Hence, the maximum excess delay is 16.7~ns. The corresponding maximum detectable path length is 5~m, which is sufficient for this small-scale measurement. Waveguides are integrated at both Tx and Rx modules, with gains around 7~dBi and half-power beamwidth (HPBW) around 60$^\circ$.

The propagation channel is confined in a 0.69~m (length) $\times$ 0.61~m (width) $\times$ 1.00~m (depth) anechoic chamber. The inside surfaces of the chamber are wrapped by absorbing materials to restrain the multi-path effect.
Tx and Rx are placed at two ends of the artificial channel.
The Tx is installed on a displacement platform, which supports horizontal and vertical movements in the $x$-$z$ plane perpendicular to the Tx-Rx line-of-sight. The displacement precision is 0.02~mm for both axes.
The Rx is static, and the center of the Tx array is aligned with the Rx, with a distance of 0.86~m.

The PC alternately controls the movement of Tx through the displacement platform and the measuring process through the VNA.
Key parameters of the measurement are summarized in Table~\ref{tab:system_parameter}.

\begin{table}
  \centering
  \caption{Parameters of the measurement system.}
    \setlength{\tabcolsep}{6mm}{
    \begin{tabular}{ll}
    \toprule
    Parameter & Value \\
    \midrule
    Frequency band              & 260-320~GHz \\
    Bandwidth                   & 60~GHz \\
    Time resolution             & 16.7~ps \\
    Space resolution            & 5~mm \\
    Sweeping interval           & 60~MHz \\
    Sweeping points             & 1001 \\
    Maximum excess delay        & 16.7~ns \\
    Maximum path length         & 5~m \\
    Waveguide gain / HPBW       & 7~dBi / 60$^\circ$ \\
    Tx-Rx distance (aligned)    & 0.86~m \\
    Displacement precision      & 0.02~mm\\
    \bottomrule
    \end{tabular}
    }
  \label{tab:system_parameter}
\end{table}

\begin{table}
  \centering
  \caption{Measurement deployment.}
    \setlength{\tabcolsep}{3mm}{
    \begin{tabular}{c|ccc}
    \toprule
    Scenario 1 & Case 1 & Case 2 & Case 3\\
    \hline
    Antenna array type & \makecell[c]{16$\times$16\\UPA} & \makecell[c]{32$\times$32\\UPA} & \makecell[c]{64$\times$64\\UPA}\\ 
    Element spacing & 0.5 mm & 0.5 mm &0.5 mm\\
    NF/FF region & FF & Boundary & NF\\
    \bottomrule
    \end{tabular}
    }
  \label{tab:deployment_parameter}
\end{table}

\subsection{Measurement Deployment}
\rev{In the measurement, as shown in Fig.~\ref{fig:platform}(b), four sides of the channel are all wrapped by absorbing materials. Three cases are measured, i.e., the virtual UPA at Tx consisting of 16$\times$16, 32$\times$32, and 64$\times$64 elements, with element spacing of 0.5~mm, as summarized in Table~\ref{tab:deployment_parameter}.}

All cases share the same center of UPA, and the Rx is perpendicularly pointing to this center.
In terms of the Rayleigh distance, arrays in Case~1, 2, and 3 belong to the far-field region, far-near-field boundary, and the near-field region, respectively.

The measurement starts from the element at the left bottom corner, and all elements in the UPA are scanned first horizontally (in the $x$-axis) and then vertically (in the $z$-axis). The measuring time for each position is about 2.3~s. However, the measuring time is subject to the motion driven by the displacement platform, which takes about 4 hours to traverse a 64$\times$64 UPA.

\section{Measurement Results} \label{section: results}
In this section, measurement results are discussed, including the properties of the line-of-sight (LoS) ray and the metal surface-reflected ray across the UPA in FF and NF cases. The variation of path gain and phase across the UPA is analyzed. First, we reveal non-linear phase change in the near-field region, and the Rayleigh distance in terms of the maximum phase error is verified. Second, the cross-field path loss model compatible with both FF and NF cases across the UPA structure is proposed.
Besides, the multi-path fading caused by the superposition of the two rays is observed. 

\subsection{Channel impulse response}

\begin{figure}
    \centering
    \includegraphics[width=\linewidth]{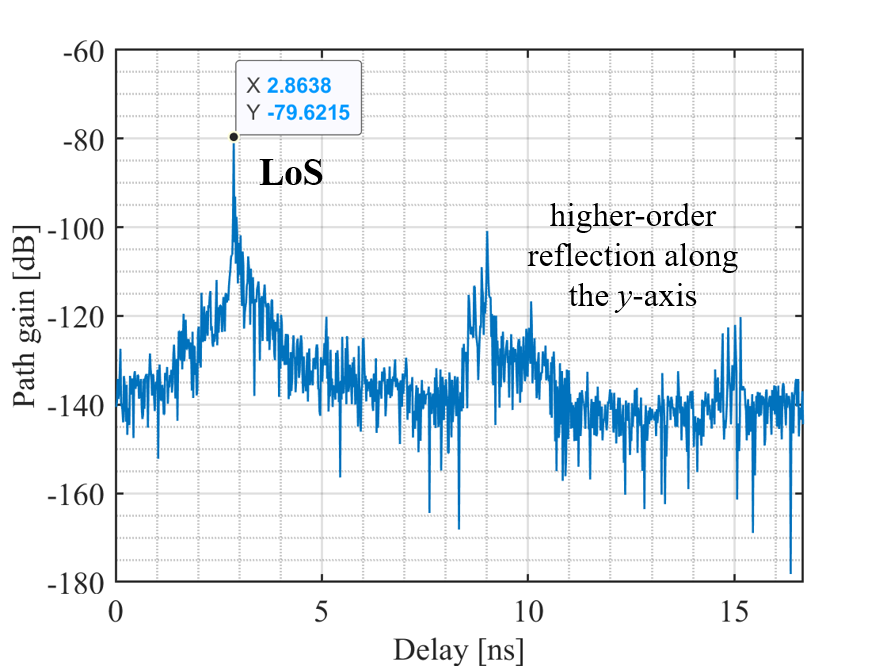}
    \caption{CIR at the UPA center.}
    \label{fig:CIR_UPA_center}
\end{figure}

\rev{In the measurement, for each channel from one element at Tx to Rx, the frequency-domain measurement derives 1001 samples in channel transfer function (CTF), with the sample interval of 60~MHz. We then transfer it to the channel impulse response (CIR) by inverse Fourier transform (IFT), with the sample interval of 16.7~ps.}
We show the CIR result from the center of the UPA to the Rx in Fig.~\ref{fig:CIR_UPA_center}. The LoS ray arrives at 2.8638~ns, corresponding to the traveling distance of 0.8591~m, and has the strongest power of 79.6~dB. The measured path loss and phase of the LoS ray across the UPA are revealed in Section~\ref{sec:pl} and Section~\ref{sec:phase}, respectively.
Since the two ends of the channel are not covered by absorbing materials, we can also observe higher-order reflected rays along the $y$-axis.

\subsection{Path loss}\label{sec:pl}

\begin{figure}
    \centering
    \includegraphics[width=\linewidth]{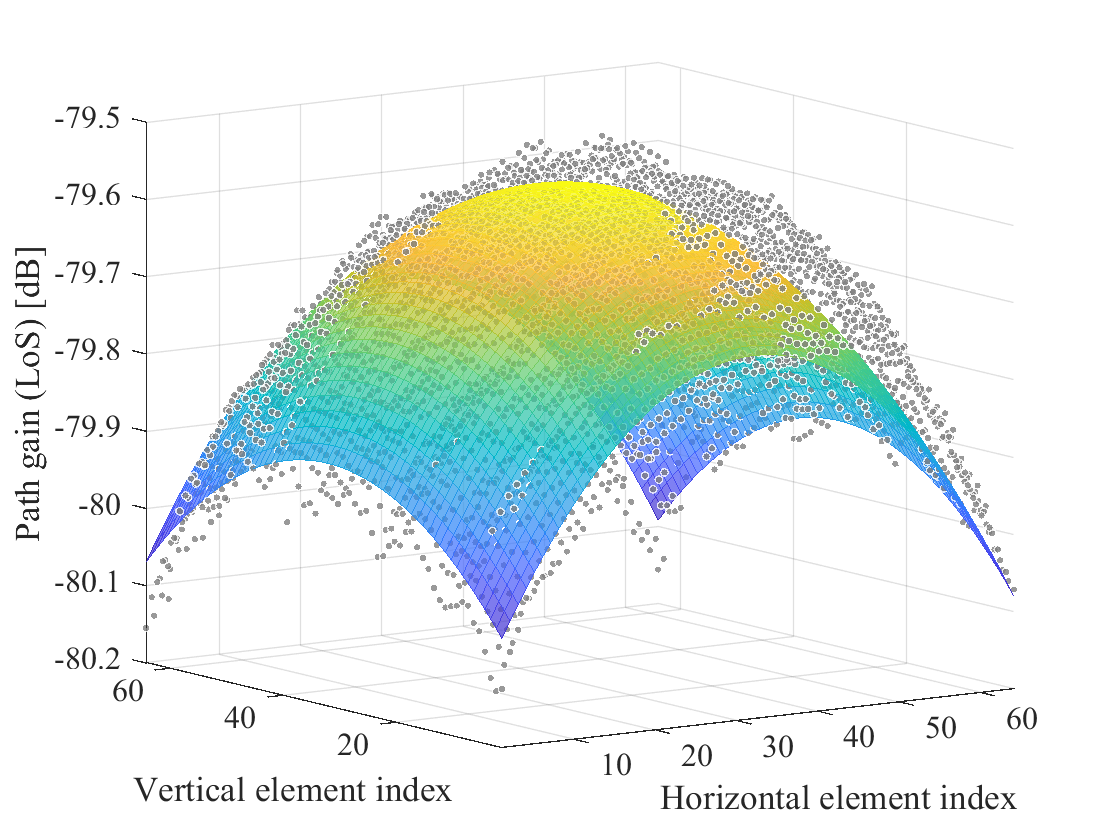}
    \caption{Path gain of the LoS ray across the UPA. Gray dots represent the measurement result, and the colored surface represents the model fitting result.}
    \label{fig:power_heatmap_64}
\end{figure}

Fig.~\ref{fig:power_heatmap_64} shows the measured path gain of the LoS ray across the $64\times64$ UPA in gray dots. 
In the far-field region, the free-space path loss (FSPL) is given by Friis’ law, which is described as
\begin{equation}\label{eq:friis}
    \text{FSPL [dB]} = 20\times\log_{10}\frac{4\pi}{\lambda}\cdot d_{\rm 3D},
\end{equation}
where $d_{\rm 3D}$ is the 3D distance between Tx and Rx, and $\lambda$ represents the carrier wavelength.
\rev{The Friis' law is derived from the transmission formula, which indicates how EM wave power is attenuated during its transmission between two antennas that are separated by a sufficiently large distance. The distance $d_{\rm 3D}$ is confined to the far-field region. Therefore, Friis’ law no longer works in the near-field region. For instance, in our experiment, the actual transmission distance ranges from 0.8600~m to 0.8603~m between the Rx and elements at Tx. According to the Friis’ law as \eqref{eq:friis}, the transmission distance corresponds to the path loss from 80.6742~dB to 80.6772~dB with variation of 0.003~dB. In contrast, the measured path loss from 79.5404~dB to 80.1555~dB with the variation of 0.6151~dB. In other words, the variation of received power at the array elements is underestimated by Friis’ Law. This is attributed to the large array size compared with the small Tx-Rx distance.}

Besides, in the measurement result, path gain reduces unequally along the $x$-axis and the $z$-axis. \rev{The explanation is as follows. In our experiment, standard rectangular WR2.8 waveguides are integrated onto the test port of two RF modules, which are used as open-ended waveguides (OEG)~\cite{kawalko1997near} with the gain around 7~dB. The gain is uniformly removed in the calibration process. However, similar to the horn antenna, whose gain decreases faster along the width (long side) than that along the height (short side)~\cite{xiao2021near}, gain reduction factors of the waveguide are also likely to be associated with the aperture structure, which accords with the measurement result that the path gain along the x-axis (long side) decreases more rapidly.}

Instead, inspired by the expression of antenna aperture gain in the NF~\cite{xiao2021near}, we propose a cross- far- and near-field path loss model for the MISO system with the UPA structure, by introducing a cross-field factor $K$, as
\begin{subequations}
\begin{align}
    &\text{PL [dB]} = 20\times\log_{10}\frac{4\pi}{\lambda}\cdot d_{\rm ref}\cdot K(\Delta x,\Delta z,\lambda,d_{0}), \\
    &K(\Delta x,\Delta z,\lambda,d_{0}) = 1 + C_{1}^{\frac{\left(\frac{\Delta x}{C_3}\right)^2 + \left(\frac{\Delta z}{C_4}\right)^2}{\lambda} - C_2\cdot d_{0} },
\end{align}
\end{subequations}
where the reference distance $d_{\rm ref}$ is a model parameter. $K$ is the cross-field factor associated with the coordinate of the element relative to the center ($\Delta x$, $\Delta z$), the wavelength $\lambda$, and the distance from the UPA center to the Rx $d_{0}$.

\rev{The proposed model provides a statistical way to characterize the power change across the UPA at a fixed Tx-Rx distance, whereas the parameters are designed to imply physical meanings. Starting at the center of the array (when $\Delta x = \Delta z = 0$), the absolute value of power is first corrected by introducing parameter $d_{\rm ref}$. Then, as we move to another position ($\Delta x$, $\Delta z$) away from the center, it turns to the question whether the movement is prominent compared with the Tx-Rx distance $d_{0}$, which is equivalent to the question whether the array at $d_{0}$ with aperture size $\sqrt{\Delta x^2+\Delta z^2}$ is in the NF region. 
Therefore, the cross-field factor $K$ is used to revise the implementation of Friis’ Law in the NF region, which increases when the NF effect becomes more significant by (i) the expansion of the UPA compared with the wavelength, (ii) the decrease of the Tx-Rx distance, and (iii) the relative degree between the above two factors.}

In the expression of the cross-field factor $K$, parameter $C_1$ is the base of the exponential function ${C_{1}}^{p}$, where the exponent $p=\left[\left(\frac{\Delta x}{C_3}\right)^2 + \left(\frac{\Delta z}{C_4}\right)^2\right] /\lambda - C_2\cdot d_{0}$ maps the obscure transition between FF and NF to the value of factor $K$. The value of $C_1$ is larger than 1 so that (i) ${C_{1}}^{p}$ and $K$ increase with $p$, i.e., when the UPA expands, the correction factor $K$ becomes more significant, and (ii) ${C_{1}}^{p}\to 0$ and $K \to 1$ as $p$ decreases. At the UPA center, $\Delta x=\Delta z=0$, and the path loss is equal to the Friis' FSPL at the equivalent distance $d_{\rm ref}\cdot(1+{C_1}^{-C_2\cdot d_0})$.

Furthermore, as the distance from the UPA center to the Rx $d_{0}$ is fixed, $C_2$ adjusts the transition between FF and NF. For instance, in this measurement, Rx is pointing perpendicular to the UPA, i.e., the spatial angle at the UPA center is $\theta$ = 0$^\circ$. In a general MISO model, the Rayleigh distance is proportional to the square of $\cos\theta$. Specifically, according to the more universal formula of the Rayleigh distance that consists of the spatial angle, $2D^2(\cos\theta)^2/\lambda$, when the UE points to the UPA center with a larger spatial angle to the UPA plane, the NF effect is less significant. Though the angle $\theta$ is not included in the model, it is already embodied in values of $C_2$, $C_3$ and $C_4$.
Besides, parameters $C_3$ and $C_4$ are intended to model the different degrees of path loss descent along two axes.

Fig.~\ref{fig:power_heatmap_64} shows the fitting result of the measured path gain across the UPA, \rev{with the mean squared error (MSE) of 0.0022.} The result is discussed as follows.
First, $d_{\rm ref} = 0.4459$, $C_1=1.3295$, $C_2=1.1433$, and the path loss at the UPA center given by the model is 79.56~dB, equivalent to the Friis' FSPL at $d_{\rm ref}\cdot(1+{C_1}^{-C_2\cdot d_0})=0.7829$~m.
Second, $C_3=0.8885$ and $C_4=1.2318$, which accords with the measurement result where the path loss decreases more rapidly along the $x$-axis, compared with the counterparts along the $z$-axis.
Thanks to the cross-field factor, the proposed path loss model is accurate across NF and FF regions.

\begin{table}
    \centering
    \caption{Phase of the LoS ray across the UPA.}
    \includegraphics[width=\linewidth]{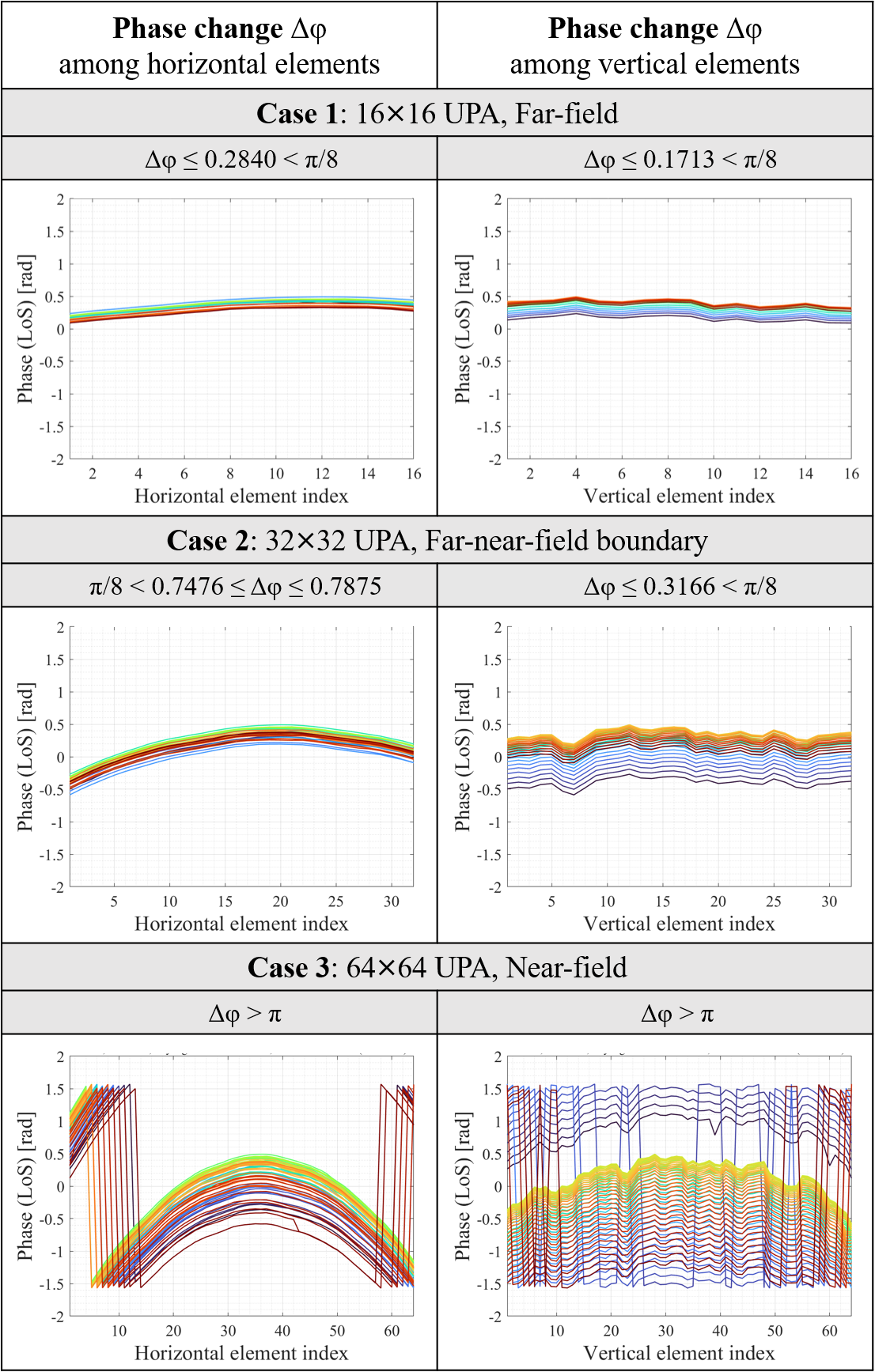}
    \label{tab:sc1_los_phase}
\end{table}

\subsection{Phase}\label{sec:phase}

Table~\ref{tab:sc1_los_phase} summarizes the phase change along horizontal ($x$-axis) and vertical ($z$-axis) positions. \rev{Colored lines indicate phase change among horizontal/vertical elements at different vertical/horizontal positions.} In the FF case, the maximum phase difference across each column and each row are both smaller than $\pi/8$, which accords with the definition of Rayleigh distance and can be regarded as a linear phase change. When the UPA expands to 32$\times$32 in Case~2 and 64$\times$64 in Case~3, the phase difference gradually exceeds $\pi/8$, and a non-linear phase change can be clearly observed. In particular, the phase change among vertical elements is less smoother, since the measurement scans the UPA first in the $x$-axis and then in the $z$-axis. In other words, the measurement at vertical elements is not continuous and is affected by the positional accuracy of the displacement platform.

\section{Conclusion} \label{section: conclusion}

In this paper, we conducted a wideband MISO channel measurement with a virtual planar array at 260-320~GHz. Channels from at most 4096 elements in the UPA are investigated. The characteristics of far- and near-field channels across the UPA are elaborated. Specifically, properties of LoS and surface-reflected rays across the UPA are analyzed, and the following observations are drawn.
First, non-linear phase change is observed in the NF region, and the Rayleigh criterion regarding the maximum phase error is verified.
Second, instead of Friis' formula, a new cross-field path loss model compatible with both FF and NF cases across the UPA structure is proposed. Specifically, a cross-field factor is parameterized in the model to reveal the transition between FF and NF cases, which is determined by the squared distance between the element and the UPA center, the carrier wavelength, and the UPA center-Rx distance.
\rev{At this stage, the proposed model is statistically accurate and has a physical meaning. The structure of the model can be generalized to other distances. Still, more measurements and further analysis at different Tx-Rx distances ($d_0$) and frequencies ($f$, or wavelength $\lambda$) are essential, in order to make a conclusion on model parameters.}




\bibliographystyle{IEEEtran}
\bibliography{bibliography}

\end{document}